\documentclass[aps,twocolumn,nofootinbib,preprintnumbers,showkeys,superscriptaddress]{revtex4-1}

\usepackage{graphicx}
\usepackage{dcolumn}
\usepackage{bm}
\usepackage{amsmath}
\usepackage{float}
\usepackage{multirow}
\usepackage{xcolor}
\usepackage{scrextend}
\usepackage[colorlinks=true, pdfstartview=FitV, bookmarks=true, bookmarksnumbered=true, breaklinks]{hyperref}
\usepackage{color}
\definecolor{blue}{rgb}{0.0, 0.0, 1.0}
\definecolor{red}{rgb}{1.0, 0.0, 0.0}
\definecolor{royalblue}{rgb}{0.0, 0.14, 0.4}
\hypersetup{linkcolor=royalblue, citecolor=blue, urlcolor=royalblue}

\def\orcid#1{\kern .08em\href{https://orcid.org/#1}{\includegraphics[keepaspectratio,width=0.7em]{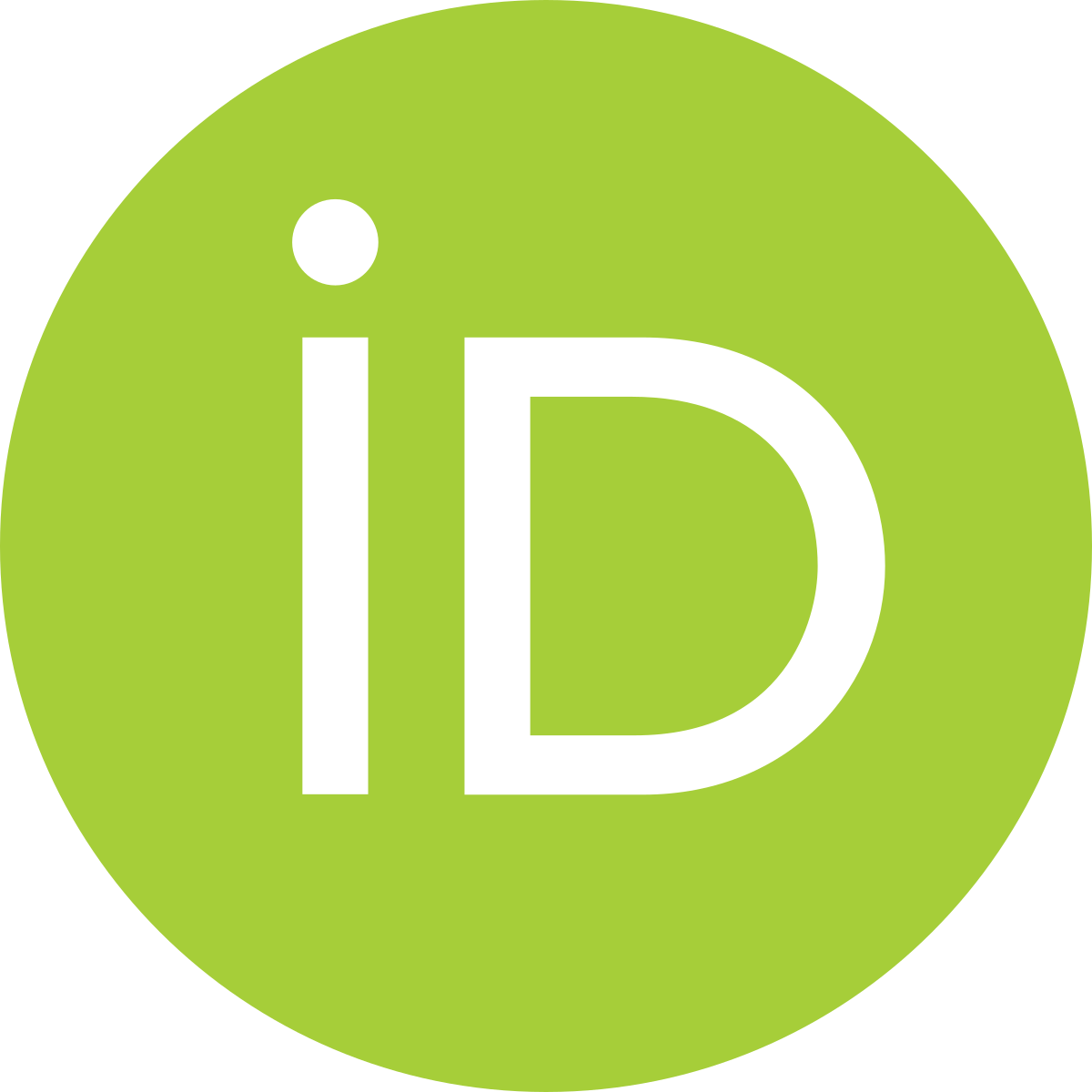}}}

\begin{document}
	
\title{Relativistic corrections to decays of heavy baryons in the quark model}

\author{Ahmad Jafar Arifi\orcid{0000-0002-9530-8993} }
\email[]{ahmad.jafar.arifi@apctp.org}
\affiliation{Research Center for Nuclear Physics (RCNP), Osaka University, Ibaraki, Osaka 567-0047, Japan}
\affiliation{Asia Pacific Center for Theoretical Physics (APCTP), Pohang, Gyeongbuk 37673, Republic of Korea}

\author{Daiki Suenaga\orcid{0000-0002-6928-4123}}
\email[]{suenaga@rcnp.osaka-u.ac.jp}
\affiliation{Research Center for Nuclear Physics (RCNP), Osaka University, Ibaraki, Osaka 567-0047, Japan}

\author{Atsushi Hosaka\orcid{0000-0003-3623-6667} }
\email[]{hosaka@rcnp.osaka-u.ac.jp}
\affiliation{Research Center for Nuclear Physics (RCNP), Osaka University, Ibaraki, Osaka 567-0047, Japan}
\affiliation{Advanced Science Research Center, Japan Atomic Energy Agency, Tokai, Ibaraki 319-1195, Japan}

\date{\today}

\begin{abstract}
We investigate relativistic corrections of an order $1/m^{2}$, where $m$ is the constituent quark mass, to heavy baryon decays by emitting one pseudoscalar meson in the quark model.
This work is motivated by shortcomings in the previous studies in the nonrelativistic 
quark model for decays of the Roper-like states such as $\Lambda_c(2765)$. 
We find that the relativistic corrections due to the internal motion of quarks 
are essential ingredients in improving their decay properties
such that the decay widths are significantly increased.
In addition, such corrections can explain a phenomenological suppression of 
the quark axial-vector coupling constant $g_A^q$ for the $\Sigma_c(2455)$ and $\Sigma_c(2520)$ decays.
\end{abstract}

\maketitle

\section{Introduction}
In recent years, many states of hadrons that contain heavy quarks are discovered 
in many experiments such as SLAC, KEK, and LHC experiments
Among them, the newly observed $\Lambda_b(6072)$ in LHC 
experiments~\cite{Sirunyan:2020gtz, Aaij:2020rkw} is particularly interesting.
From its mass and decay properties, this state is suggested to be 
the first radial excitation of the $\Lambda_b$ baryon~\cite{Arifi:2020yfp,Azizi:2020ljx},
 which is an analogous state of the Roper resonance, $N(1440)$~\cite{Roper:1964zza}.
In fact, there are other candidates of the Roper-like states with 
heavy quark flavors such as $\Lambda_c(2765)$ and $\Xi_c(2970)$.
Such analogy is further confirmed from the recent result by Belle 
that $\Xi_c(2970)$ favors spin-parity $1/2^+$~\cite{Moon:2020gsg}.
Interestingly, they appear to have not only a similar excitation 
energy but also a large decay width~\cite{Arifi:2020yfp}.

Historically, the Roper resonance has been a mysterious state 
because the observed mass is much lower than that predicted by the quark model, 
and the inverse ordering with the negative parity state is puzzling. 
These unexpected results have induced great amount of discussions to 
understand its nature both experimentally~\cite{Zyla:2020zbs,Aznauryan:2009mx} and theoretically~\cite{JuliaDiaz:2006av,Santopinto:2004hw,Giannini:2015zia,Takayama:1999kc,Glozman:1995fu,Brown:1983ib,Gutsche:2017lyu,Brodsky:1997de}.
One promising physical interpretation is that the Roper 
resonance is a quark core coupled by meson clouds~\cite{Suzuki:2009nj,Burkert:2017djo}.
Moreover, the relativistic effects are found to be 
important~\cite{Kubota:1976ft,Capstick:1994ne,Capstick:1986bm}.

One of the difficulties of the Roper-like states is that their observed 
broad decay widths seem to contradict with the prediction of the narrow width 
by the quark model~\cite{Nagahiro:2016nsx, Arifi:2020ezz,Zhong:2007gp,Liu:2012sj,Guo:2019ytq,Liang:2020kvn}.
Due to a large discrepancy, one may expect that the nonresonant 
process or $f_0(500)$ contribution may be essential.
However, in our previous study,
it is suggested that such a contribution is insignificant~\cite{Arifi:2020yfp}.
The suppression is also supported by the chiral effective model~\cite{Suenaga:2021qri}.
Therefore, the prediction of the narrow width is indeed a serious 
problem of the previous studies in the quark model.

In this work, we investigate relativistic corrections in the constituent quark model 
primarily to solve the problem in the decays of the Roper-like states.  
At the same time, we also study the corrections to other low-lying states.
For this purpose, we will use the Foldy-Wouthuysen-Tani (FWT) transformation.
This method was employed long ago by Kubota and Ohta in analyzing 
the photoexcitation amplitudes of nucleon resonances~\cite{Kubota:1976ft}.
They emphasized that such corrections are crucial to give the correct 
sign of the photoexcitation amplitude of $N(1440)$, 
leading to better agreement with the data.
The relativistic effects in the photoexcitation amplitudes are also 
confirmed by other computation~\cite{Capstick:1994ne}.
Furthermore, the relativistic treatments give better agreement for the 
mass of $N(1440)$~\cite{Capstick:1986bm}.
Motivated by these observations, we expect that the relativistic 
corrections will also play important roles for heavy baryons.

\begin{figure}[b]
	\centering
	\includegraphics[scale=0.2]{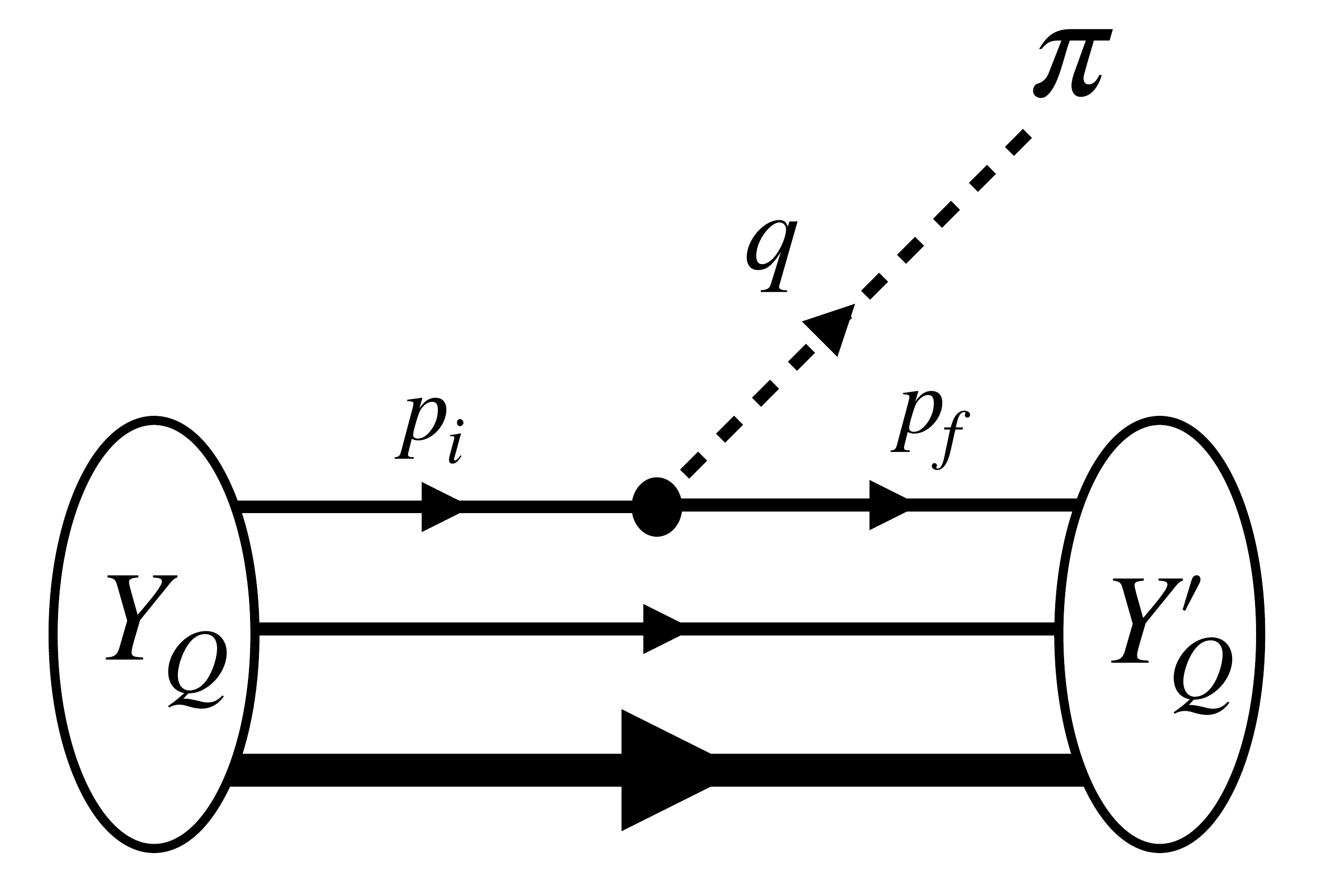}
	\caption{\label{decay} 
		Schematic picture of one-pion emission decay of heavy baryon 
		$Y_Q$ in the quark model, where the pion is regarded as a Nambu-Goldstone boson. }
\end{figure}

\begin{table*}[t]
	\begin{ruledtabular}
		\caption{ Decay widths estimated by the nonrelativistic quark model 
			with the relativistic corrections (NR+RC) and without them (NR) for 
			various charmed baryons in units of MeV. The minimum and maximum 
			values of the decay widths are found numerically within the parameter 
			range as indicated in the section of {\it model parameters}.}
		\centering
		\begin{tabular}{lcccccc}
			State			 		 	  			& Multiplet																	      & 	Channel						 & $\Gamma_{\rm{NR}}$  	 &  $\Gamma_{\rm{NR + RC}}$	 	& $\Gamma_{\rm{Exp.}}$ & Ref \\ \hline
			
			$\Sigma_c(2455)^{++}$  &$\Sigma_c(1S, 1/2(1)^+)$										  &  	$\Lambda_c\pi$		  &  4.27 - 4.34 						& 0.36 - 1.95 			 & $1.84\pm 0.04$  &Belle~\cite{Lee:2014htd}	\\ \vspace{0.2cm}
			$\Sigma_c(2520)^{++}$  &$\Sigma_c(1S, 3/2(1)^+)$										  &  	 $\Lambda_c\pi$	   	   & 29.8 - 31.4 				   		& 2.70 - 14.1			  & $14.77\pm 0.25$ 	& Belle~\cite{Lee:2014htd}	  \\  \vspace{0.2cm}
			$\Lambda_c(2595)^+$	   &$\Lambda_c(1P_\lambda, 1/2(1)^-)$						&  	$\Sigma_c(2455)\pi$	   & 1.35 - 3.16 						& 1.36 - 3.20 			 & $2.6\pm 0.6$			& CDF~\cite{Aaltonen:2011sf}			 \\
			$\Lambda_c(2625)^+$	   &$\Lambda_c(1P_\lambda, 3/2(1)^-)$						&  	 $\Sigma_c(2455)\pi$	& 0.08 - 0.15 				 & 	0.01 - 0.06 	 		& 	\\ 
			&																						&  	 $\Sigma_c(2520)\pi$ & 0.07 - 0.18 				  		    &  0.08 - 0.20 			  &  \\ \vspace{0.2cm}
			&																						&  	Sum	  							  & 0.15 - 0.33 						&	0.09 - 0.26			&$<0.97$				& CDF~\cite{Aaltonen:2011sf}	\\ 			
			$\Lambda_c(2765)^+$	  	&$\Lambda_c(2S_{\lambda\lambda}, 1/2(0)^+)$	   &  	$\Sigma_c(2455)\pi$	     & 0.71 - 2.66  				  &  5.56 - 26.1  			   & 	 \\ 
			&																						 &  $\Sigma_c(2520)\pi$ &  0.67 - 2.04 					 & 5.26 - 22.9				 &  \\ \vspace{0.2cm}
			&																						 &  Sum	  						    & 1.38 - 4.70 					&	10.8 - 49.0			& $73 \pm 5$		& Belle~\cite{Abe:2006rz}			\\ 			
			$\Lambda_c(3136)^+$~\footnote{\label{rho}Masses are estimated in the quark model for the $\rho$-mode Roper-like state.} 	  &$\Lambda_c(2S_{\rho\rho}, 1/2(0)^+)$	   			  		&  	$\Sigma_c(2455)\pi$	     & 2.22 - 42.0  		 &  106 - 657  			   & 	 \\ 
														&																						 &  	$\Sigma_c(2520)\pi$	  &  8.78 - 81.3 		 & 208 - 1142				 &  \\ \vspace{0.2cm}
														&																						 &  	Sum	  							  & 11.0 - 123	 		 & 	314 - 1799			& 			 \dots & \dots 		\\ 	\hline	\vspace{0.2cm} 
			
			$\Xi_c^{'+}$					&$\Xi'_c(1S, 1/2(1)^+)$											  &  $\Xi_c\pi$	 				  & \dots~\footnote{The null results for the decay width are due to insufficient phase space.}			& \dots 				& \dots 	 		& \dots 	 \\ \vspace{0.2cm}
			$\Xi_c(2645)^+$					 &$\Xi'_c(1S, 3/2(1)^+)$												  &  	$\Xi_c\pi$	   						& 5.16 - 5.26 					& 0.93 - 2.75					& $2.06 \pm 0.13$ & Belle~\cite{Yelton:2016fqw}\\ \vspace{0.2cm}
			$\Xi_c(2790)^+$					 &$\Xi_c(1P_\lambda, 1/2(1)^-)$								  &  	$\Xi_c'\pi$	  						 & 4.24 - 11.6 					& 4.33 - 11.7		 			& $8.9 \pm 1.0$	& Belle~\cite{Yelton:2016fqw}\\
			$\Xi_c(2815)^+$					 &$\Xi_c(1P_\lambda, 3/2(1)^-)$								  &  	$\Xi'_c\pi$	   						& 0.18 - 0.34 					& 0.04 - 0.12					 &    \\ 
			&																					&  	$\Xi_c(2645)\pi$	  		  & 1.83 - 4.16 					& 1.26 - 2.92				    	&  \\  \vspace{0.2cm}
			&																					&  	Sum	  										 	& 2.01 - 4.50 					& 1.30 - 3.04					& $2.43 \pm 0.26$			& Belle~\cite{Yelton:2016fqw}\\ 			
			$\Xi_c(2970)^+$					 &$\Xi_c(2S_{\lambda\lambda}, 1/2(0)^+)$			 	&  	 $\Xi'_c\pi$		  				  & 0.16 - 0.87 			& 1.62 - 8.36					&   \\  
			&																					&  	 $\Xi_c(2645)\pi$		  		 & 0.23 - 0.95 					& 2.07 - 9.40						 & 	  \\ 
			&																					&  	 $\Sigma_c(2455) K$		  	& 0.16 - 0.35 					& 0.38 - 1.18						&  		\\  \vspace{0.2cm}
			&																					&  	Sum	  									 & 0.55 - 2.17 					&	4.07 - 18.9						& $28.1\pm 2.4$ 	 & Belle~\cite{Yelton:2016fqw}\\ 
			$\Xi_c(3318)^+$~\footref{rho}	&$\Xi_c(2S_{\rho\rho}, 1/2(0)^+)$	&  	 $\Xi'_c\pi$		  			& 0.00 - 7.58 			  & 17.6 - 133					  		&   \\  
										  &															  &  	 $\Xi_c(2645)\pi$		  & 0.35 - 13.1 			& 37.4 - 238						 & 	  \\  
										 &															  &  	 $\Sigma_c(2455) K$	& 2.61 - 24.5 			   & 29.6 - 183						&  		\\ 
										  &															  &  	Sum	  							 & 2.96 - 45.2 				&	84.6 - 554						 &  \dots & \dots 	\\		
		\end{tabular}
		\label{result1}
	\end{ruledtabular}
\end{table*}

\begin{table*}[t]
	\begin{ruledtabular}
		\caption{Similar to Table.~\ref{result1}, but for bottom baryons. }
		\centering
		\begin{tabular}{lcccccc}
			State			 		 	  			& Multiplet																	      & 	Channel						 & $\Gamma_{\rm{NR}}$  	 &  $\Gamma_{\rm{NR + RC}}$	 	& $\Gamma_{\rm{Exp.}}$ & Ref \\ \hline
			$\Sigma_b(5810)^{+}$	&$\Sigma_b(1S, 1/2(1)^+)$										   &  	$\Lambda_b\pi$			 & 11.9 - 12.3 		& 0.62 - 5.11			& $4.83 \pm 0.31$ & LHCb~\cite{Aaij:2018tnn} \\ \vspace{0.2cm}
			$\Sigma_b(5830)^{+}$	&$\Sigma_b(1S, 3/2(1)^+)$										   &  	$\Lambda_b \pi$		 	 & 20.4 - 21.4 		& 1.08 - 8.80			& $9.34 \pm 0.47$ & LHCb~\cite{Aaij:2018tnn} \\ \vspace{0.2cm}
			$\Lambda_b(5912)^{0}$ &$\Lambda_b(1P_\lambda, 1/2(1)^-)$					   &  	$\Sigma_b \pi$				   & 0.001 - 0.003	& 0.001 - 0.003		 & $<0.25$ & LHCb~\cite{Aaij:2020rkw}	\\ \vspace{0.2cm}
			$\Lambda_b(5920)^{0}$ &$\Lambda_b(1P_\lambda, 3/2(1)^-)$					   &  	$\Sigma_b^{*}\pi$			& 0.004 - 0.008	& 0.004 - 0.009		&  $<0.19$ & LHCb~\cite{Aaij:2020rkw}\\ 
			$\Lambda_b(6072)^{0}$ &$\Lambda_b(2S_{\lambda\lambda}, 1/2(0)^+)$	 &  	$\Sigma_b \pi$	   			& 0.72 - 2.17 		& 4.97 - 20.8		  & 	 \\ 
			&																						&  	$\Sigma_b^{*}\pi$	   & 1.08 - 3.00 		& 7.81 - 31.5		  & 	  \\ \vspace{0.2cm}
			&																						&  	Sum	  							 & 1.80 - 5.17 		   &	12.8 - 52.3		& $72\pm 11$ & LHCb~\cite{Aaij:2020rkw}	\\ 
			$\Lambda_b(6469)^{0}$~\footnote{\label{rho2}Masses are estimated in the quark model for the $\rho$-mode Roper-like state.} &$\Lambda_b(2S_{\rho\rho}, 1/2(0)^+)$		 			   &  	$\Sigma_b \pi$	    & 3.29 - 51.6 	 & 107 - 725		&  \\ 
														&																						&  	$\Sigma_b^{*}\pi$  & 8.05 - 102 	& 215 - 1396	   &  \\ \vspace{0.2cm}
														&																						&  	Sum	  						 & 11.3 - 154 	  &	322 - 2121		 & 	 \dots &  \dots  \\  \hline \vspace{0.2cm}
			$\Xi_b(5935)^{-}$			  	&$\Xi'_b(1S, 1/2(1)^+)$						&   $\Xi_b\pi$		 	 & 0.25 - 0.25 			& 0.04 - 0.13			& $<0.08$ & LHCb~\cite{Aaij:2014yka}\\ \vspace{0.2cm}
			$\Xi_b(5945)^{-}$			 	&$\Xi'_b(1S, 3/2(1)^+)$						&  	 $\Xi_b\pi$		  	 & 2.87 - 2.90  		& 0.43 - 1.43			& $1.65 \pm 0.31$ & LHCb~\cite{Aaij:2014yka}\\ \vspace{0.2cm}
			$\Xi_b(6096)^-$~\footnote{\label{missing}Masses are taken from Ref~\cite{Chen:2018orb}.}					 &$\Xi_b(1P_\lambda, 1/2(1)^-)$		 &  	$\Xi_b'\pi$	  	  & 2.40 - 5.49 		 & 2.41 - 5.50		 	 & \dots & \dots\\
			$\Xi_b(6100)^-$~\footnote{Mass is taken from the latest result~\cite{Sirunyan:2021vxz}.} 					 &$\Xi_b(1P_\lambda, 3/2(1)^-)$		 &  	$\Xi'_b\pi$	   	  & 0.01 - 0.01 		 & 0.00 - 0.00			 &    \\ 
			&														  &  	$\Xi_b^{*}\pi$	 & 0.91 - 1.93 			& 0.63 - 1.36			&  \\  \vspace{0.2cm}
			&														  &  	Sum	  				  & 0.92 - 1.94 		 & 0.63 - 1.36		  	 & $<1.9$					& CMS~\cite{Sirunyan:2021vxz} \\ 			
			$\Xi_b(6255)^-$~\footref{missing} 					 &$\Xi_b(2S_{\lambda\lambda}, 1/2(0)^+)$	 &   $\Xi'_b\pi$		 & 0.19 - 0.72 		  		& 1.49 - 6.33		&    \\  
			&																				&  	$\Xi_b^{*}\pi$	  & 0.31 - 1.13 			& 2.75 - 11.4		 & 	 \\ 	\vspace{0.2cm}
			&																				&  	Sum	  				   & 0.50 - 1.85 			 &	4.24 - 17.7		  & \dots	 & \dots  \\ 
			$\Xi_b(6647)^-$~\footref{rho2} 				&$\Xi_b(2S_{\rho\rho}, 1/2(0)^+)$						&  	$\Xi'_b\pi$		  	& 0.14 - 8.80 			   & 17.9 - 132		&   \\ 
														 &																				&  	$\Xi^{*}_b\pi$	  & 0.28 - 14.7 			& 36.3 - 261		 & 	 \\  
														 &																				&  	Sum	  				   & 0.42 - 23.5 			 &	54.2 - 393		  &     \dots & \dots \\
		\end{tabular}
		\label{result2}
	\end{ruledtabular}
\end{table*}

\section{Nonrelativistic quark model}
The quark model computation of heavy baryon decays follows Refs.~\cite{Nagahiro:2016nsx,Arifi:2020ezz}.
The harmonic oscillator wave functions of baryons are formed in the heavy quark basis. 
They are denoted as $Y_Q(nl_\xi, J(j)^P)$, where $nl$ stand for 
the node and orbital angular momentum quantum numbers, 
and $\xi =\lambda$ or $\rho$ indicate the two internal excitation modes of quarks.
Its spin $(J)$ and parity $(P)$ together with the brown muck spin $(j)$ are denoted by $J(j)^P$.

In the quark model, the one-pion emission decay of heavy baryon is depicted in Fig.~\ref{decay}.
Here, we employ the axial-vector type coupling for the interaction 
between the pion and a light quark inside a heavy baryon as
\begin{eqnarray}
\mathcal{L}_{\pi qq} = \frac{g_A^q}{2f_\pi} \bar{q} \gamma_\mu\gamma_5 \vec{\tau} q \cdot \partial^\mu \vec{\pi}. \label{inter}
\end{eqnarray}
This interaction is inspired by the low-energy theorem of chiral symmetry.
In many cases, the nonrelativistic calculations have been performed 
by expanding the interaction in powers of $1/m$ and considering the terms up to order $1/m$ as given by
\begin{eqnarray}
H_{NR} = g  \left[ \boldsymbol{\sigma}\cdot \boldsymbol{q} + \frac{\omega_\pi}{2m}\left( \boldsymbol{\sigma}\cdot \boldsymbol{q} -2 \boldsymbol{\sigma}\cdot \boldsymbol{p}_i \right) \right], \label{nonrel}
\end{eqnarray}
where we define $g=g_A^q/2f_\pi$ with $g^q_A=1$ the quark axial-vector 
coupling constant and $f_\pi=93$ MeV the pion decay constant.
Here we denote the energy and momentum of the outgoing pion as $(\omega_\pi, \boldsymbol{q})$.
For kaon emission decays, the parameters such as the kaon decay 
constant ($f_K=111$ MeV), energy and momentum should be changed accordingly.
The initial and final momenta of the light quark are denoted by $\boldsymbol{p}_i$ and $\boldsymbol{p}_f$.

In the previous works~\cite{Zhong:2007gp,Nagahiro:2016nsx, Arifi:2020ezz}, decay widths of heavy baryons were investigated by using the interaction in Eq.~(\ref{nonrel}).
However, the resulting decay widths turned out to be too small for the Roper-like states, e.g. 
$\Lambda_c(2765)$ baryonas shown in the column denoted as $\Gamma_{\rm{NR}}$ of Table~\ref{result1}.

\section{Relativistic corrections of order $\boldsymbol{1/m^2}$}
To estimate them properly, we perform the FWT transformation~\cite{Greiner:1990tz} for the Lagrangian in Eq~(\ref{decay}). 
After some calculations, we obtain
\begin{eqnarray}
H_{RC}  = \frac{g}{8 m^2}\biggl[m_\pi^2\boldsymbol{\sigma}\cdot \boldsymbol{q} +  2 \boldsymbol{\sigma} \cdot ( \boldsymbol{q} -2\boldsymbol{p}_i ) \times (\boldsymbol{q} \times \boldsymbol{p}_i)  \biggr],\quad \label{relcor}
\end{eqnarray}
where $m_\pi$ is the pion mass.
Note that $m_\pi$ should be replaced by $m_K$ for the kaon emission decay.
What we found in this work is that the term proportional to $\boldsymbol{p}_i^2$ 
in the second term of Eq.~(\ref{relcor}) plays an important role not only 
for the Roper-like state but also for $\Sigma_c$'s.
This term is due to the internal motion of the quarks inside a heavy baryon.
In the electromagnetic interaction, such a term appears as the spin-orbit coupling 
in the relativistic correction~\cite{Kubota:1976ft}.

\section{Model parameters}
In the quark model, there are three parameters: the light quark mass $m$, 
the heavy quark mass $M$, and the spring constant $k$.
Following our previous study~\cite{Nagahiro:2016nsx}, 
for $\Lambda_c$ and $\Lambda_b$ baryons we will use the constituent quark masses as
$m_{u(d)} = 0.35\pm 0.05\ {\rm GeV},$ 
$M_c = 1.5\pm 0.1\ {\rm GeV},$ and
$M_b = 5.0\pm 0.1\ {\rm GeV}$.
For $\Xi_c$ and $\Xi_b$ baryons they consist of three different quarks so that 
we use the averaged mass $m = 0.40\pm 0.05\ {\rm GeV}$ for the $u, d,$ and $s$ quarks. 
Here we have used the strange quark mass as $m_s = 0.45\pm 0.05\ {\rm GeV}$.
In this work, the spring constant is adjusted as $k = 0.03 \pm 0.01\ {\rm GeV^3}$ in order to
get the level spacing around $\omega_\lambda=\sqrt{k(2m+M)/(mM)}=0.35 \pm 0.05$ GeV which is the typical excitation energy of the first excited state of heavy baryons. 
We will use the same value of the spring constant for various quark flavors contents.
From the above parameters, we obtain the range parameter of the harmonic oscillator wave functions as $a_\lambda =\sqrt{m_\lambda \omega_\lambda}= 0.40 \pm 0.04\ {\rm GeV}$ and $a_\rho= \sqrt{m_\rho \omega_\rho} = 0.29\pm 0.03\ {\rm GeV}$ for $\Lambda_c$ baryons where we define $m_\lambda=2mM/(2m+M)$, $m_\rho=m/2$, and $\omega_\rho=\sqrt{3k/m}$. 
The values of the range parameters slightly vary for $\Xi_c, \Lambda_b$ and $\Xi_b$.

In actual computations, we use the heavy baryon masses from the experimental data when available. Otherwise, we use the theoretical input as given in the footnotes of Tables~\ref{result1} and ~\ref{result2}.
In the following, we will look at the decays of low-lying states one by one. 

\section{Ground states}
Let us start from the $\Sigma_c(2455)$ and $\Sigma_c(2520)$. 
These states are regarded as ground states because the quarks are in the lowest $S$-wave orbit.  
However, they have an energy excess due to the spin-one (bad) diquark 
that can decay into the spin-zero (good) diquark by emitting one pion.  

The nonrelativistic quark model of order $1/m$ overpredicts the 
decay widths of $\Sigma_c$ states and their siblings by a factor of two as shown in the column denoted 
as $\Gamma_{\rm{NR}}$ of Tables~\ref{result1} and ~\ref{result2}.
In our previous study~\cite{Nagahiro:2016nsx}, the discrepancy has 
led to the discussion of the suppression factor of about 3/4 for the quark axial-vector coupling constant $g^q_A$.
In the literature, the universal suppression parameter is introduced 
to explain the experimental data~\cite{Zhong:2007gp}.
The necessity of the suppression factor for $g_A^q$ has been known for long time 
for the nucleon $g_A$; in the nonrelativistic quark model $g_A = 5/3$, about 30\% larger 
than the observed value $g_A \sim 1.25$~\cite{Yamaguchi:2019vea}.  
The situation is essentially the same for the decay of $\Sigma_c \to \Lambda_c \pi$.

Now, let us see the suppression mechanism by including the relativistic corrections in more detail.
The matrix element of the leading term of order $1/m^0$ 
is the spin-isospin factor of $\sigma_i \tau_a$ times 
the overlap of the common ground state wave functions for $\Sigma_c$ and  $\Lambda_c$
which is unity in the long-wavelength limit of the pion momentum.
For the term of order $1/m$, the matrix elements of 
$\boldsymbol{\sigma}\cdot \boldsymbol{p}_i$ and $\boldsymbol{\sigma}\cdot \boldsymbol{q}$ 
cancel each other giving only a small contribution of around $0.1\%$ 
of the total width by using the interaction in Eq.~(\ref{nonrel}). 
The cancellation can be understood since the ratio $R_{p/q}=\left<\boldsymbol{\sigma}\cdot \boldsymbol{p}_i\right>/\left<\boldsymbol{\sigma}\cdot \boldsymbol{q}\right>$ is around 0.42 for this case.
In the relativistic corrections of order $1/m^2$, 
the matrix element of $\boldsymbol{p}_i^2$ in Eq.~(\ref{relcor})
gives a factor proportional to the square of the range parameter $a^2$.
This term appears with the opposite sign to the leading term of $1/m^0$. 
This explains the reduction of the quark axial-vector coupling constant $g_A^q$.
As shown in Tables~\ref{result1} and ~\ref{result2}, 
it is fair to say that the agreement with the data is improved 
when observing that the data marginally fall into the calculated range. 

\section{Negative parity states}
These are the first excited states of quark orbital motion in the $P$ wave ($1P$ state).
We expect that they are dominated by the lower $\lambda$ modes.
We assume that this is the case not only for the $1P$ states,
but also for the $2S$ Roper-like states in the following.

The relativistic correction is found to be insignificant for the negative parity states.
For instance, the correction to the decay of $\Lambda_c(2595)$ with $J^P=1/2^-$ is 
negligible and the interaction in Eq.~(\ref{nonrel}) is sufficiently good in explaining the experimental data.
For this decay, the leading term of order $1/m^0$ with $\boldsymbol{\sigma}\cdot \boldsymbol{q}$ is negligible 
because it results in a term proportional to $q^2$ which is vanishing in the long-wavelength limit.
Meanwhile, for the term of order $1/m$, the matrix element of $\boldsymbol{\sigma}\cdot \boldsymbol{p}_i$ gives a finite term of order $q^0$. 
As a result, the $\boldsymbol{\sigma}\cdot \boldsymbol{p}_i$ becomes the dominant term.
This is in the line with the $S$-wave decay of $\Lambda_c(2595)\to\Sigma_c(2455)\pi$.
For the relativistic correction terms of order $1/m^2$ as in Eq.~(\ref{relcor}), 
the matrix element is found to give only a term proportional to $q^2$ resulting in a small contribution.
For the case of $\Lambda_c(2625)$ with $J^P=3/2^-$, 
the relativistic correction is found to be sizable for $\Sigma_c(2455)\pi$ channel. 
However, because of the $D$-wave nature, 
the actual value is relatively small and the agreement with the data is still good.
This behavior applies to other siblings such as $\Xi_c(2790)$ 
and $\Xi_c(2815)$ as given in Tables~\ref{result1} and ~\ref{result2}.

Very recently, CMS collaboration observed the new $\Xi_b(6100)$ state with a narrow width $\Gamma < 1.9$ MeV~\cite{Sirunyan:2021vxz}. 
It is found that our prediction of the decay width agrees well with the data as shown in Table~\ref{result2}, when the spin and parity are identified as $3/2^-$.

\section{Roper-like states}
Now, let us come to the main result of the present work. 
Here we found that the relativistic correction is essential for the Roper-like states.
As discussed earlier, the nonrelativistic quark model predicts narrow widths around 
a few MeV that are smaller than the experimental data by one order of magnitude.
However, by taking into account the relativistic corrections in Eq.~(\ref{relcor}), 
the decay widths are significantly improved and have better agreement with 
the data as shown in the column denoted as $\Gamma_{\rm{NR + RC}}$ of Tables~\ref{result1} and ~\ref{result2}.

It is also worth mentioning that there are other decay modes, e.g., $f_0(500)$ contribution in a two-pion emission decay, that may contribute to the total width of the Roper-like state. However, from the experimental observations, such a contribution is insignificant as discussed in our previous studies~\cite{Arifi:2020ezz,Arifi:2020yfp}.

The shortcoming in the nonrelativistic quark model 
can be understood from the orthogonality of the wave functions.
The leading term of order $1/m^0$ with $\boldsymbol{\sigma}\cdot \boldsymbol{q}$, 
which is the spin-flip transition process, contains a vanishing overlap of the 
orthogonal orbital wave functions in the long-wavelength limit.
In contrast, the $\boldsymbol{\sigma}\cdot \boldsymbol{p}_i$ term 
of order $1/m$  in Eq.~(\ref{nonrel}) provides a finite contribution.
However, the odd power of the quark momentum operator will translate 
into the pion momentum $q$ and always come with the pion energy $\omega_\pi$, 
which makes the role of $\boldsymbol{\sigma}\cdot \boldsymbol{p}_i$ term 
not very important resulting in only small decay widths up to order $1/m$~\cite{Nagahiro:2016nsx}.
On the other hand, in the relativistic correction of $1/m^2$, the matrix elements 
consist of the higher terms of the quark momentum of $\boldsymbol{p}^2_i$ as given in Eq.~(\ref{relcor}).
The even power of the quark momentum operator will translate into 
the square of the range parameter $a^2$ giving considerable contributions.
Together with the $\boldsymbol{\sigma}\cdot \boldsymbol{p}_i$ term of order $1/m$ with the same sign, the corrections of order $1/m^2$ lead to a large increase of the total decay widths.

From the above discussion, it is essential to include the next leading order term ($1/m^2$ term) especially when the leading term is suppressed. 
In other words, the $1/m^2$ term is the leading term for the decay of the Roper-like states. 
In contrast to other cases such as the ground states and the negative parity states, the leading order term play the dominant role.

As anticipated earlier, we have so far discussed the $\lambda$-mode excited states.
To complete our discussions, we also mention the results for $\rho$-mode ones.
The excitation energies of the $\rho$ modes are expected to be larger;
in the harmonic oscillator base, we expect that the mass of the $\rho$-mode Roper-like state is about 1 GeV above the ground state.
In more realistic calculations with a linear confinement potential, 
this energy is somewhat lowered~\cite{Yoshida:2015tia}.
We expect that the mass of the $\rho$-mode Roper-like state is about 
850 MeV above the ground state $\Lambda_c(2286)$.
In Tables~\ref{result1} and \ref{result2}, results are shown by using this value.
The resulting widths are largely increased.
Note that there are also other possible decay modes such as $D$ meson emission decay, that make the width even larger.
Therefore, we consider that this could be the reason that the $\rho$-mode Roper-like state is not likely to be observed.

For the $\Lambda_c(2765)$ and $\Lambda_b(6072)$ baryons, 
the computed decay widths are found to be similar.
This behavior follows the heavy-quark flavor symmetry~\cite{Manohar:2007}; i.e., 
the dynamics of charmed and bottom baryons are similar.
Also, the branching ratio $R=\Gamma(\Sigma_c(2520)\pi)/\Gamma(\Sigma_c(2455)\pi)$ 
is not significantly changed with the inclusion of the relativistic corrections 
and still consistent with the prediction from the heavy-quark spin symmetry~\cite{Isgur:1991wq}.
For the case of Roper-like $\Xi_{c(b)}$ baryons, the decay widths are found to 
be smaller than Roper-like $\Lambda_{c(b)}$ baryons despite having a similar phase space. 
This can be understood by the fact that the $\Xi_{c(b)}$ baryons have only one light quark that couples to a pion. 

For the case of $\Xi_c$, the $\Sigma_c(2455)K$ channel is open. 
In this case, the relativistic correction for the kaon emission decay is not large as 
compared to the pion emission decay because of the smaller phase space volume. 
As a result, the ratio of $\Sigma_c(2455)^{++}K^-$ to $\Xi_c(2645)^0\pi^+$ becomes 
smaller around 10\% when the relativistic correction is included as compared to the case without it, which is around 40\%.
This prediction can be tested in the experiment to further clarify the role of relativistic effects for the Roper-like states.

\section{Summary}
We have investigated relativistic corrections up to order $1/m^2$ to the decays of 
low-lying heavy baryons through pseudoscalar meson emission in the quark model.
As a result, we have found that the agreement with the data is significantly improved. 
In particular, the decay widths of the $\Lambda_c(2765)$ and other Roper-like states 
are greatly increased by one order of magnitude as compared to the previously calculated values up to order $1/m$. Our present work implies that a better relativistic approach is desired in analyzing baryon decays, which is a challenging problem.

It is emphasized that we do not need a suppression of the quark axial-vector 
coupling constant $g_A^q$ by hand~\cite{Yan:1992gz}, 
but rather it is naturally explained by the relativistic effect.
The fact that we can consistently use $g_A^q=1$ supports the discussion by 
Weinberg on the mended symmetry for the quark axial-vector coupling constant~\cite{Weinberg:1990xn,Weinberg:1990xm}.

\section*{Acknowledgements}
A. J. A thanks Research Center for Nuclear Physics (RCNP) for the hospitality during his stay in completion of this work. A. J. A is also supported by the YST Program at the APCTP through the Science and Technology Promotion Fund and Lottery Fund of the Korean Government and also by the Korean Local Governments - Gyeongsangbuk-do Province and Pohang City.
We also thank Kiyoshi Tanida for useful discussions. A. H. is supported in part by Grants-in Aid for Scientific Research, Grant No. 17K05441(C) and by Grants-in Aid for Scientific Research on Innovative Areas (Grant No. 18H05407).

\end{document}